\documentclass[nofootinbib,twocolumn,prd,superscriptaddress,preprintnumbers,floatfix]{revtex4-1}
\usepackage{slashed,bm,graphicx,color,upgreek}
\usepackage{amssymb}
\usepackage{mathrsfs}
\usepackage{amsmath}
\usepackage[normalem]{ulem}
\usepackage{bbm}
\RequirePackage[colorlinks=true,urlcolor=blue,anchorcolor=blue,citecolor=blue,filecolor=blue,
linkcolor=blue,menucolor=blue,linktocpage=true,pdfproducer=medialab]{hyperref}

\newcommand{\blue}[1]{\color{blue} #1 \color{black} }

\newcommand{\unit}[1]{\mathrm{\,#1}}
\newcommand{\T}{\mathbf{T}}
\newcommand{\U}{\mathbf{U}}
\newcommand{\Bs}{B}
\newcommand{\cO}{\mathcal{O}}
\newcommand{\cR}{\mathcal{R}}
\newcommand{\cF}{\mathcal{F}}
\newcommand{\GeV}{\,\mathrm{GeV}}

\newcommand{\beq}{\begin{equation}}
\newcommand{\eeq}{\end{equation}}

\begin{document}

\title{Baryon Non-Invariant Couplings in Higgs Effective Field Theory}%

\author{Luca Merlo}
\author{Sara Saa}
\author{Mario Sacrist\'an-Barbero}
\affiliation{\vspace{1mm} 
Departamento de F\'isica Te\'orica and Instituto de F\'sica Te\'orica, IFT-UAM/CSIC, Universidad Aut\'onoma de Madrid, Cantoblanco, 28049, Madrid, Spain}

\begin{abstract}
The basis of leading operators which are not invariant under baryon number is constructed within the Higgs Effective Field Theory. This list contains 12 dimension six operators, which preserve the combination $B-L$, to be compared to only 6 operators for the Standard Model Effective Field Theory. The discussion on the independent flavour contractions is presented in detail for a generic number of fermion families adopting the Hilbert series technique.
\end{abstract}

\date{\today}

\preprint{\blue{FTUAM-16-45}}
\preprint{\blue{IFT-UAM/CSIC-16-135}}

\maketitle

%
%
\section{Introduction}

The Standard Model (SM) cannot explain the present matter-antimatter asymmetry in our universe~\cite{Gavela:1993ts,Gavela:1994ds,Gavela:1994dt}. A possibility to tackle this problem is to consider additional sources of baryon number violation, as predicted in several Beyond the SM (BSM) contexts, such as Grand Unified Theories~\cite{Georgi:1974sy}. On the other side, no baryon number ($B$) violating (BNV) process has been observed so far, despite the numerous experimental searches on BNV decays of nucleons -- which provide the most stringent constraints -- hadrons, heavy quarks and leptons, and $Z$ boson~\cite{Olive:2016xmw}.

Without assuming any specific model, an effective field theory (EFT) approach can be adopted to describe BNV processes. The first attempt in this direction goes back to the late 1970's~\cite{Weinberg:1979sa,Wilczek:1979hc,Weldon:1980gi,Abbott:1980zj}, followed by a few more recent studies~\cite{Buchmuller:1985jz,Grzadkowski:2010es,Alonso:2014zka}. All these analyses are performed in the so-called SM Effective Field Theory (SMEFT) context, characterised by the construction of non-renormalisable operators, invariant under the SM gauge symmetries, and built up in terms of SM fermions, gauge bosons and the $SU(2)_L$-doublet scalar boson (``Higgs'' for short)~\cite{Englert:1964et,Higgs:1964ia,Higgs:1964pj}. The cut-off of the theory suppressing these operators will be referred to as $\Lambda_{\Bs}$. At the lowest order in the expansion in $1/\Lambda_{\Bs}$, four BNV independent structures of canonical dimension $d=6$ were identified~\cite{Weinberg:1979sa,Wilczek:1979hc,Weldon:1980gi,Abbott:1980zj},
\begin{equation}
\begin{aligned}
\cO_{1}&=\bar{d}^{C}_{R\alpha }u_{R\beta } \ \bar{Q}^{C}_{L\gamma i}L_{Lj}  \ \epsilon_{ij}\ \epsilon_{\alpha\beta\gamma}\,,\\
\cO_{2}&=\bar{Q}^{C}_{Li\alpha }Q_{Lj\beta } \ \bar{u}^{C}_{R\gamma }e_{R } \ \epsilon_{ij}\ \epsilon_{\alpha\beta\gamma}\,,\\
\cO_{3}&=\bar{Q}^{C}_{Li\alpha }Q_{Lj\beta } \ \bar{Q}^{C}_{L\gamma k}L_{Ll} \ \epsilon_{il}\,\epsilon_{kj}\ \epsilon_{\alpha\beta\gamma}\,,\\
\cO_{4}&=\bar{d}^{C}_{R\alpha }u_{R\beta } \ \bar{u}^{C}_{R \gamma }e_{R} \ \epsilon_{\alpha\beta\gamma}\,,
\end{aligned}
\label{SMEFT1}
\end{equation}
where $Q_L\equiv(u_L, d_L)^T$, $u_R$, $d_R$, $L_L\equiv(\nu_L, e_L)^T$, and $e_R$ are the SM fermions, $\epsilon_{\alpha\beta\gamma}$ and  $\epsilon_{ij}$ are the antisymmetric tensors for the colour and electroweak (EW) contractions.
If right-handed (RH) neutrinos, $N_R$, are considered in addition, this set is extended by two operators:
\begin{equation}
\begin{aligned}
\cO_{5}&=\bar{Q}^{C}_{Li\alpha }Q_{Lj\beta } \ \bar{d}^{C}_{R\gamma }N_{R } \ \epsilon_{ij} \epsilon_{\alpha\beta\gamma}\,, \\
\cO_{6}&=\bar{u}^{C}_{R\alpha }d_{R\beta } \ \bar{d}^{C}_{R\gamma }N_{R} \ \epsilon_{\alpha\beta\gamma},
\end{aligned}
\label{SMEFT2}
\end{equation}
The operators listed in the previous equations refer only to one generation of fermions. Moving to the three generation case does not require the introduction of additional structures, but only to insert explicitly flavour indices on the fermion fields. 

The operators in Eqs.~(\ref{SMEFT1}) and (\ref{SMEFT2}) preserve $B-L$ with $\Delta B=+1=\Delta L$, and then a baryon can only decay into an anti-lepton and a meson. The constraints on the proton lifetime~\cite{Nishino:2009aa,Nishino:2012bnw,Miura:2016krn} translate into a lower bound on the cut-off $\Lambda_{\Bs}$ of about $10^{15}\unit{GeV}$, independently of the specific flavour contraction that can be considered for each operator. On the contrary, when a flavour symmetry is considered, such as the so-called Minimal Flavour Violation ansatz in its global~\cite{Chivukula:1987py,D'Ambrosio:2002ex,Cirigliano:2005ck,Davidson:2006bd,Alonso:2011jd,Feldmann:2010yp,Alonso:2011yg,Alonso:2012fy,Alonso:2013mca,Alonso:2013nca} or gauged~\cite{Grinstein:2010ve,Feldmann:2010yp,Guadagnoli:2011id,Buras:2011zb,Buras:2011wi,Feldmann:2016hvo,Alonso:2016onw} versions, the scale $\Lambda_{\Bs}$ can be lowered, but still it will be much larger than the electroweak scale $v\approx246\unit{GeV}$.

The basic ingredient of the SMEFT construction is the treatment of the Higgs field as an exact EW doublet. Although this hypothesis is currently supported by collider searches (see for example Ref.~\cite{Butter:2016cvz}), the present uncertainties leave open the possibility for alternative descriptions of the EW symmetry breaking (EWSB) mechanism, potentially free from the Hierarchy problem. Still in the context of effective approaches, a description that allows for deviations from the exact EW doublet representation for the Higgs field is the so-called Higgs Effective Field Theory (HEFT) Lagrangian that generalises the SMEFT one. The HEFT Lagrangian is the most general description of gauge and Higgs couplings, respecting the paradigm of Lorentz and $SU(3)_c\times SU(2)_L\times U(1)_Y$ gauge invariance: it is a very useful tool to describe an extended class of ``Higgs'' models, from the SM and the SMEFT scenarios, to Goldstone Boson Higgs models~\cite{Kaplan:1983fs,Kaplan:1983sm,Banks:1984gj,Agashe:2004rs,Gripaios:2009pe,Feruglio:2016zvt,Gavela:2016vte} and dilaton-like constructions~\cite{Goldberger:2008zz,Goldberger:2008zz,Vecchi:2010gj,Matsuzaki:2012mk,Chacko:2012vm,Bellazzini:2012vz}. 

The aim of this paper is to construct the BNV operator basis in the HEFT context, completing in this way previous studies on the HEFT framework.  

In the next section, the HEFT setup is summarised and the BNV basis is presented. The comparison between the HEFT basis and the corresponding one in the SMEFT setup is discussed in Sect.~\ref{Sect:Comparison}. The counting of the distinct flavour contractions, considering a generic number of fermions, is performed in Sect.~\ref{Sect:Hilbert}, based on the Hilbert series technique. The latter is a mathematical method from Invariant Theory to count the number of independent structures invariant under a certain symmetry group (for recent phenomenology applications see Refs.~\cite{Jenkins:2009dy,Hanany:2010vu,Lehman:2015via,Henning:2015daa,Lehman:2015coa,Henning:2015alf}).

%
%
\section{The BNV HEFT Lagrangian}
\label{Sect:BNV}

The crucial difference between the SMEFT and the HEFT is the relationship between the physical Higgs field $h(x)$ and the SM Goldstone bosons (GBs) $\overrightarrow{\pi}(x)$: in the SMEFT, the four fields belong to the $SU(2)_L$ doublet $\Phi(x)$,
\begin{equation}
\Phi(x)=\U(x)\begin{pmatrix} 0 \\ \frac{v+h(x)}{\sqrt{2}} \end{pmatrix}\,,
\label{HiggsEq}
\end{equation}
where
\beq
\U(x)\equiv e^{i\overrightarrow\sigma\cdot\overrightarrow\pi(x)/v}
\eeq
is the GB matrix. In the HEFT, instead, the physical Higgs and the GB matrix are treated as independent objects~\cite{Feruglio:1992wf,Contino:2010mh,Alonso:2012px,Alonso:2012pz,Buchalla:2013rka,Brivio:2014pfa,Gavela:2014vra,Gavela:2014uta,Brivio:2016fzo}. This fact, together with the adimensionality of the GB matrix, leads to a much larger number of operators in the HEFT with respect to the SMEFT, at the same order in the expansion. As a consequence, HEFT exhibits the following distinguishing features~\cite{Alonso:2012jc,Brivio:2013pma,Alonso:2014wta,Hierro:2015nna,Brivio:2016fzo,Brivio:2015kia,Eboli:2016kko}:
\begin{itemize}
\item[-] several correlations typical of the SMEFT, such as those between triple and quartic gauge couplings, are lost in the HEFT;
\item[-] Higgs couplings are completely free in the HEFT, while they can be correlated to pure gauge couplings in the SMEFT;
\item[-] some couplings that are expected to be strongly suppressed in the SMEFT, are instead predicted with higher strength in the HEFT and are potentially visible in the present LHC run.
\end{itemize}

In Ref.~\cite{Brivio:2016fzo}, the complete HEFT Lagrangian, invariant under baryon and lepton numbers, has been presented at first order in the expansion on the new physics scale~\cite{Gavela:2016bzc}, making explicit the custodial symmetry nature of the operators. The building blocks used for these structures are the SM gauge bosons, the GB matrix $\U(x)$, the physical Higgs field $h(x)$ and the SM fermions arranged in doublets of the global $SU(2)_L$ or $SU(2)_R$ symmetries. Arranging the RH fermions\footnote{RH neutrinos are considered as part of the $SU(2)_R$ lepton doublet, but the origin of their masses will not be discussed here.} in doublets of $SU(2)_R$,
\beq
Q_{R}=\begin{pmatrix} u_{R} \\ d_{R} \end{pmatrix} \qquad 
L_{R}=\begin{pmatrix} N_{R} \\ e_{R} \end{pmatrix}\,,
\label{doublets}
\eeq
allows to distinguish the custodial symmetry preserving operators from those that instead violate it. Furthermore, this notation is consistent with the HEFT formalism, where the GBs matrix $\U(x)$ transforms as a bi-doublet of the global $SU(2)_L\times SU(2)_R$ symmetry,
\beq
\U(x)\rightarrow L\, \U(x)\, R^\dag\,,
\label{Umatrix}
\eeq
being $L$, $R$ the unitary transformation associated to $SU(2)_{L,R}$, respectively. Indeed, the Yukawa couplings are given by
\beq
\bar{Q}_{L}\U(x)Y_{Q}Q_{R}\,,\qquad\qquad \bar{L}_{L}\U(x)Y_{L}L_{R}\,,
\eeq
where the Yukawa matrices are written in a compact notation as $6\times 6$ block-diagonal matrices in the flavour space, $Y_{Q}={\rm diag}(Y_{u},Y_{d})$ and $Y_{L}={\rm diag}(Y_{\nu},Y_{e})$. Furthermore, it is useful to introduce the scalar chiral field $\T(x)$,
\beq
\T(x)\equiv \U(x)\sigma_3 \U(x)^\dag\,,\qquad
\T(x)\rightarrow L \T(x) L^\dag\,,
\label{Tmatrix}
\eeq
that breaks $SU(2)_R$, while preserving $SU(2)_L$, and therefore behaves as a spurion for the custodial symmetry.

In the HEFT Lagrangian, the dependence on the physical Higgs is conventionally described through adimensional generic functions $\cF(h/v)$~\cite{Feruglio:1992wf,Grinstein:2007iv}, being $v$ the EW vacuum expectation value. These functions are commonly written as a polynomial expansion in $h/v$, $\cF(h/v)=1+\alpha (h/v)+\beta (h/v)^2+\ldots$, which follows from the fact that the physical Higgs is an isosinglet scalar of the EW symmetry.
The study of the scalar field manifold, depending on the specific $\cF(h)$, can indeed lead to phenomenological consequences, allowing to disentangle between different frameworks. This has been analysed in Refs.~\cite{Alonso:2015fsp,Alonso:2016btr,Alonso:2016oah}.\\

One could expect that the basis of BNV operators introduced in Eqs.~(\ref{SMEFT1}) and (\ref{SMEFT2}) will not be modified in the HEFT framework, as they are purely fermionic. Indeed, these six operators are simply rewritten in terms of $SU(2)_L$ and $SU(2)_R$ fermion doublets. However, the fact that the GB matrix $\U$ and the chiral scalar field $\T$ are adimensional allows to construct additional independent structures with the same canonical dimensions. 

The set of operators that constitutes the BNV HEFT basis, at the first order in the expansion on $\Lambda_{\Bs}$, consists of 12 independent structures: 
\beq
\begin{aligned}
\cR_{1}&=\bar{Q}^{C}_{Li\alpha }Q_{Lj\beta } \, \bar{Q}^{C}_{Lk\gamma }L_{Ll} \, \epsilon_{il}\, \epsilon_{kj} \, \epsilon_{\alpha\beta\gamma} \, \mathcal{F}_{1}(h)\\
\cR_{2}&=\bar{Q}^{C}_{Li\alpha }Q_{Lj\beta } \ \bar{Q}^{C}_{Lk\gamma }(\T L_{L})_{l} \, \epsilon_{il}\,\epsilon_{kj} \, \epsilon_{\alpha\beta\gamma} \, \mathcal{F}_{2}(h)\\[3mm]
\cR_{3}&=\bar{Q}^{C}_{Ri\alpha }Q_{Rj\beta } \ \bar{Q}^{C}_{Rk\gamma }L_{Rl} \, \epsilon_{il}\,\epsilon_{kj} \, \epsilon_{\alpha\beta\gamma} \, \mathcal{F}_{3}(h)\\
\cR_{4}&=\bar{Q}^{C}_{Ri\alpha }Q_{Rj\beta } \, \bar{Q}^{C}_{Rk\gamma }(\U ^{\dagger}\T \U L_{R})_{l} \, \epsilon_{il}\,\epsilon_{kj} \, \epsilon_{\alpha\beta\gamma} \, \mathcal{F}_{4}(h)\\[3mm]
\cR_{5}&=\bar{Q}^{C}_{Ri\alpha }Q_{Rj\beta }\bar{Q}^{C}_{Lk\gamma }L_{Ll} \, \epsilon_{ij}\,\epsilon_{kl} \, \epsilon_{\alpha\beta\gamma} \, \mathcal{F}_{5}(h)\\
\cR_{6}&=\bar{Q}^{C}_{Ri\alpha }Q_{Rj\beta }\bar{Q}^{C}_{Lk\gamma }(\T L_{L})_{l} \, \epsilon_{ij}\,\epsilon_{kl} \ \epsilon_{\alpha\beta\gamma} \, \mathcal{F}_{6}(h)\\
\cR_{7}&=(\bar{Q}^{C}_{R\alpha }\U ^{t})_{i}(\T \U Q_{R\beta })_{j}\bar{Q}^{C}_{Lk\gamma }L_{Ll} \, \epsilon_{il}\,\epsilon_{kj} \, \epsilon_{\alpha\beta\gamma} \, \mathcal{F}_{7}(h)\\
\cR_{8}&=(\bar{Q}^{C}_{R\alpha }\U ^{t})_{i}(\T \U Q_{R\beta })_{j}\bar{Q}^{C}_{Lk\gamma }(\T L_{L})_{l} \, \epsilon_{il}\,\epsilon_{kj} \, \epsilon_{\alpha\beta\gamma} \, \mathcal{F}_{8}(h)\\[3mm]
\cR_{9}&=\bar{Q}^{C}_{Li\alpha }Q_{Lj\beta }\bar{Q}^{C}_{Rk\gamma }L_{Rl} \, \epsilon_{ij}\,\epsilon_{kl} \, \epsilon_{\alpha\beta\gamma} \, \mathcal{F}_{9}(h)\\
\cR_{10}&=\bar{Q}^{C}_{Li\alpha }Q_{Lj\beta }\bar{Q}^{C}_{Rk\gamma }(\U ^{\dagger}\T \U L_{R})_{l} \, \epsilon_{ij}\,\epsilon_{kl} \, \epsilon_{\alpha\beta\gamma} \, \mathcal{F}_{10}(h)\\
\cR_{11}&=(\bar{Q}^{C}_{L\alpha }\U^*)_i(\U^\dagger\T Q_{L\beta })_ j\bar{Q}^{C}_{Rk\gamma }L_{Rl} \, \epsilon_{il}\,\epsilon_{kj} \, \epsilon_{\alpha\beta\gamma} \, \mathcal{F}_{11}(h)\\
\cR_{12}&=(\bar{Q}^{C}_{L\alpha }\U^*)_i(\U^\dagger\T Q_{L\beta })_ j\bar{Q}^{C}_{Rk\gamma }(\U ^{\dagger}\T \U L_{R})_{l} \, \epsilon_{il}\,\epsilon_{kj}\, \epsilon_{\alpha\beta\gamma} \, \mathcal{F}_{12}(h)\,.
\end{aligned}
\label{HEFT1}
\eeq
Other BNV operators can be constructed, but are redundant with respect to the structures in this list. For example,  one could consider an operator similar to $\cR_2$, with $\T$ contracted to the second quark doublet instead than to the lepton doublet: however, $\cR_1$, $\cR_2$ and this alternative operator are not independent among each other and one should choose only two of them. Other examples will be discussed in Sect.~\ref{Sect:Hilbert}.

All the operators in this list have canonical mass dimension $6$ and therefore are suppressed by $\Lambda^2_{\Bs}$. Indeed, the insertion of the scalar chiral field $\T$ or of the GB matrix does not lead to any additional mass suppression. Among these 12 operators, only 4 of them are custodial symmetry preserving, $\cR_{1}$, $\cR_{3}$, $\cR_{5}$ and $\cR_{9}$, and thus do not contain the custodial spurion $\T$.

When ignoring RH neutrinos, the number of independent operators reduces to 9: in particular, $\cR_4$, $\cR_{10}$ and $\cR_{12}$ turn out to be vanishing or redundant with respect to the other structures.

\section{Comparison with the SMEFT}
\label{Sect:Comparison}

The BNV SMEFT operators in Eqs.~(\ref{SMEFT1}) and (\ref{SMEFT2}) and the ones in Eq.~(\ref{HEFT1}) present a series of similarities:
\begin{enumerate}
\item[-] all the operators can be written in terms of scalar currents, being the other type of contractions vanishing or redundant by Fierz identity; 
\item[-] both bases contain operators classified into four distinct classes: schematically, $Q_{L}Q_{L}Q_{L}L_{L}$, $Q_{R}Q_{R}Q_{R}L_{R}$, $Q_{L}Q_{L}Q_{R}L_{R}$ and $Q_{R}Q_{R}Q_{L}L_{L}$;
\item[-] the operators in both bases preserve $B-L$.
\end{enumerate}
On the other side, there is not a one-to-one relation between the two sets of operators, as indeed:
\begin{enumerate}
\item[-] the $d=6$ SMEFT basis consists of only 6 independent operators, while the HEFT one presents 12 structures;
\item[-] only two combinations of SMEFT operators, $\cO_{4}-\cO_{6}$ and $\cO_{2}+\cO_{5}$, in Eqs.~(\ref{SMEFT1}) and (\ref{SMEFT2}), contain sources of custodial symmetry breaking; on the other hand, all the operators in Eq.~(\ref{HEFT1}) are custodial symmetry breaking, except for $\cR_{1}$, $\cR_{3}$, $\cR_{5}$ and $\cR_{9}$;
\item[-] $B-L$ non-invariant operators can be found in the SMEFT Lagrangian at dimensions different from six~\cite{Weinberg:1980bf}, while this is not the case in the HEFT, where indeed $B-L$ invariance is guaranteed by hypercharge invariance. This follows from two facts: first, hypercharge can be identified with $B-L$ in theories invariant under the $SU(2)_L\times SU(2)_R$ symmetry, such as in left-right symmetric models~\cite{Pati:1974yy,Mohapatra:1974gc}. In these frameworks, as the RH fermions also belong to an $SU(2)$ doublet representation, and they have the same electric charge as their left-handed (LH) counterparts, both LH and RH fields must have the same hypercharge, $-1$ for leptons and $1/3$ for quarks, in a given convention. In a compact notation, then, hypercharge can be written as $B-L$:
\beq
\begin{aligned}
\psi_L&\to e^{i(B-L)\theta(x)}\psi_L\\
\psi_R&\to e^{i(B-L)\theta(x)}e^{i\theta(x)\sigma_3}\psi_R\,,\\
\end{aligned}
\eeq
where $\theta(x)$ is the transformation parameter. The second fact which guarantees the identification of hypercharge and $B-L$ is that the only spurion breaking $SU(2)_R$, in the HEFT context is the scalar chiral field $\T$. As it does not carry hypercharge, its insertion in an operator cannot lead to hypercharge violation, neither of $B-L$. 

In the SMEFT, where hypercharge and $B-L$ are independent, SM gauge invariant operators can violate $B-L$, and the lowest dimensional example is the so-called Weinberg operator $(\bar{L}^c_L\tilde{\Phi}^*)(\tilde\Phi^\dag L_L)$. In HEFT, this operator cannot be constructed, unless other sources of $SU(2)_R$ violation are considered. As a title of example, one could consider the Pauli matrix $\sigma_+=(\sigma_1+i\sigma_2)/2$, that allows to write the equivalent to the Weinberg operator in HEFT~\cite{Hirn:2005fr}:
\beq
(\bar{L}^c_L\U^*)\sigma_+(\U^\dag L_L)\,.
\eeq
This operator preserves hypercharge, but violates $SU(2)_R$ and lepton number by two units, as it can be seen by writing explicitly the transformation under hypercharge of the GB matrix:
\beq
\U(x)\to \U(x) e^{-i\theta(x)\sigma_3}\,.
\eeq 
Notice that this is a three dimensional operator and therefore provides a direct mass term for the light active neutrinos. In contrast, the Weinberg operator in the SMEFT is of $d=5$ and thus suppressed by a power of the mass scale at which lepton number is broken. This is an example of the strong impact of the adimensionality of the GB matrix $\U$ with respect to the $SU(2)_L$ doublet Higgs of the SMEFT. In the rest of the paper, no other sources of $SU(2)_R$ violation will be considered beside $\T$, consistently with previous studies in the HEFT context.
\end{enumerate}

It is interesting to determine the connection between the operators in Eq.~(\ref{HEFT1}) and those in Eqs.~(\ref{SMEFT1}) and (\ref{SMEFT2}), as it will help to identify possible ways to distinguish the two descriptions. 
The connection for the HEFT operators which do not contain GBs is straightforward:
\beq
\begin{aligned}
\cR_1
&\rightarrow
\cO_3\\
\cR_3
&\rightarrow
\cO_4+\cO_6\\
\cR_4
&\rightarrow
-\cO_4+\cO_6\\
\cR_5
&\rightarrow
-\cO_1\\
\cR_9
&\rightarrow
\cO_2-\cO_5\\
\cR_{10}
&\rightarrow
-\cO_2-\cO_5\,.
\end{aligned}
\label{linear} 
\eeq
Notice, indeed, that the combination $\U ^{\dagger}\T \U$ appearing in $R_4$, $R_{10}$ and $R_{12}$ simplifies to $\sigma_3$ once using the definition of $\T$ in Eq.~(\ref{Tmatrix}).
This list shows that there is a linear correspondence between 6 operators of the HEFT basis and the 6 operators of the $d=6$ SMEFT one. The other HEFT operators contain interactions that can be described by SMEFT operators with dimension 8. An example is the following:
\beq
\cR_2
\rightarrow
\bar{Q}^{C}_{Li\alpha}Q_{Lj\beta}\bar{Q}^{C}_{Lk\gamma}\left[\left(\tilde\Phi\tilde\Phi^\dag-\Phi\Phi^\dag\right)L_{L}\right]_l \epsilon_{il}\epsilon_{kj} \ \epsilon_{\alpha\beta\gamma}
\eeq
where the $h$-independent couplings of the combination $\tilde\Phi\tilde\Phi^\dag-\Phi\Phi^\dag$ in the unitary gauge play the same role as the scalar chiral field $\T$ in $\cR_2$. 

The study of the connections between the HEFT and SMEFT operators leads to the conclusion that several correlations typical of the SMEFT are lost in the HEFT and that some couplings that are expected to be strongly suppressed in the SMEFT are instead predicted to be relevant in HEFT. This fact has already been pointed out in Refs.~\cite{Brivio:2013pma,Gavela:2014vra,Brivio:2016fzo} for the $B$ and $L$ invariant couplings and is confirmed here for the $B$ and $L$ non-invariant ones.
An example is the comparison between the decay rates of the proton and of the neutron: $\Gamma(p\to \pi^0e^+)$ and $\Gamma(n\to\pi^0\bar{\nu}_e)$. In the $d=6$ SMEFT framework, the values of these two observables are predicted to be exactly the same, while this correlation can be broken considering $d=8$ operators. On the other side, in the HEFT context, the operators $R_2$, $R_6$, $R_7$, $R_8$, $R_{11}$, $R_{12}$ contribute differently to the two decay rates, and no correlation arises at any order. An experimental discrepancy among these two observables could then be explained either in terms of the SMEFT, but advocating $d=8$ contributions, or in terms of the HEFT Lagrangian. The magnitude of the discrepancy is what could tell which is the correct description: a relative difference between the two decay rates larger than about $\left(v^2/\Lambda_{\Bs}^2\right)^2$ cannot be compatible with the $d=8$ SMEFT Lagrangian, and instead could well be accounted for in the HEFT context.

At present, the non-observation of the proton decay puts a lower bound on the ratio $\Lambda_{\Bs}/c_i$ of about $10^{15}\GeV$, where $c_i$ represents the combination of the operator coefficients entering the proton decay rate. 
As a result, this strategy to disentangle the two frameworks is an interesting feature from the theoretical side, although experimentally is not viable yet. Moreover, it allows to estimate the order of magnitude of the contributions to these decay rates from the $d=8$ SMEFT operators of about $10^{-51}$, with respect to those from the $d=6$ ones.

%
%
\section{Flavour contraction counting}
\label{Sect:Hilbert}

The number of independent flavour contractions can be counted directly considering the symmetries of the operators in Eq.~\eqref{HEFT1}. Alternatively, one can adopt the Hilbert series technique, which provides a polynomial function whose terms can be matched with the operators in Eq.~\eqref{HEFT1} and the corresponding coefficients count the number of independent flavour contractions. Although the matching is straightforward in the absence of scalar fields, as for the BNV HEFT operators considered here, one should be careful when dealing with structures containing the fields $\T$ and $\U$, in order to remove the redundancies due to $\T^2 =\mathbbm{1}$ and $\U^\dag \U =\mathbbm{1}$.

The discussion on the number of flavour contractions adopting the Hilbert series technique is presented below, considering in all generality $N_f$ fermion families.

The counting for $\cR_1$ is $N_f^2(2N_f^2+1)/3$ and coincides with the one in Ref.~\cite{Alonso:2014zka}, where it is discussed in terms of flavour representations by using Young tableaux. The counting of $\cR_2$ is the same as $\cR_1$, as $\T$ only adds a flip of sign in the second component of the lepton doublet. A few cases with $\T$ insertions in the $Q_{L}Q_{L}Q_{L}L_{L}$ ($LLLL$ for brevity) operators are redundant and have been subtracted from the total counting. 

For the $Q_{R}Q_{R}Q_{R}L_{R}$ ($RRRR$) operators, $\cR_3$ and $\cR_4$, which are written exclusively in terms of $SU(2)_R$ doublets, the counting simply mirrors that of the $LLLL$ ones and each operator presents $N_f^2(2N_f^2+1)/3$ flavour contractions. This is not consistent with the results in the SMEFT case (see Refs.~\cite{Alonso:2014zka,Liao:2016qyd}), where the total number of flavour contractions for the $RRRR$ structures, $\cO_4$ and $\cO_6$, is $2N_f^4$. This apparent contradiction is easily solved noticing that the $SU(2)_R$ symmetry is still partially preserved in the operators $\cR_3$ and $\cR_4$ and prevents part of the possible flavour contractions among four RH singlet fermions. Indeed, rewriting explicitly the flavour indices $a,b,c,d$, one gets
\beq
\begin{aligned}
\cR_3^{abcd}
&=
\cO_4^{\{bc\}ad}+\cO_6^{\{bc\}ad}\,,\\
\cR_4^{abcd}
&=
-\cO_4^{\{bc\}ad}+\cO_6^{\{bc\}ad}\,,
\end{aligned}
\eeq
where the brackets should be read as $\cO_i^{\{ab\}cd}\equiv \cO_i^{abcd} +\cO_i^{bacd}$. This shows that $\cR_3$ and $\cR_4$ only contain the flavour symmetric contractions in $b$ and $c$ of the SMEFT operators.
The flavour antisymmetric contractions are instead described by two additional structures: 
\beq
\begin{aligned}
\cR_3^\prime&=\bar{Q}^{C}_{Ri\alpha }(\U ^{\dagger}\T \U Q_{R\beta })_j \ \bar{Q}^{C}_{Rk\gamma }L_{Rl} \, \epsilon_{il}\,\epsilon_{kj} \, \epsilon_{\alpha\beta\gamma}\,,\\
\cR_4^\prime&=\bar{Q}^{C}_{Ri\alpha }(\U ^{\dagger}\T \U Q_{R\beta })_j \, \bar{Q}^{C}_{Rk\gamma }(\U ^{\dagger}\T \U L_{R})_{l} \, \epsilon_{il}\,\epsilon_{kj} \, \epsilon_{\alpha\beta\gamma}\,.
\end{aligned}
\eeq
These two operators are redundant with respect to $R_3$ and $R_4$ for $N_f=1$, but they should be added to the list in Eq.~(\ref{HEFT1}) for $N_f>1$ (see Ref.~\cite{Abbott:1980zj} for a similar discussion in the SMEFT). The number of the flavour contractions of these four $RRRR$ operators sums up to $2N_f^4$ matching the result for the SMEFT case.

The operators $\cR_5$--$\cR_8$ exhibit a $Q_{R}Q_{R}Q_{L}L_{L}$ ($RRLL$) structure. Among these, only $\cR_5$ can be directly related to a $d=6$ operator of the SMEFT Lagrangian. Rewriting the expression for $\cR_5$ in Eq.~(\ref{linear}), making explicit the flavour indices, one can see that $\cR_5$ only contains part of the interactions described by $\cO_1$:
\beq
\cR_5^{abcd}=-\cO_1^{\{ab\}cd}\,.
\eeq
Similarly, the operator $\cR_6^{abcd}$ contains only the flavour contractions symmetric in $a$ and $b$. It is therefore necessary to introduce two additional operators that completely break the $SU(2)_R$ structure between the first two $SU(2)_R$ quark doublets in $\cR_5$ and $\cR_6$:
\beq
\begin{aligned}
\cR^\prime_{5}&=\bar{Q}^{C}_{Ri\alpha }\left(\U ^{\dagger}\T \U Q_{R\beta }\right)_j\bar{Q}^{C}_{Lk\gamma }L_{Ll} \, \epsilon_{ij}\,\epsilon_{kl} \, \epsilon_{\alpha\beta\gamma} \, \mathcal{F}_{5}(h)\,.\\
\cR^\prime_{6}&=\bar{Q}^{C}_{Ri\alpha }\left(\U ^{\dagger}\T \U Q_{R\beta }\right)_j\bar{Q}^{C}_{Lk\gamma }(\T L_{L})_{l} \, \epsilon_{ij}\,\epsilon_{kl} \ \epsilon_{\alpha\beta\gamma} \, \mathcal{F}_{6}(h)\,.\\
\end{aligned}
\eeq
As for the previous case, these two structures are redundant with $\cR_5$ and $\cR_6$ for $N_f=1$, otherwise they should be added to the basis. $\cR^\prime_{5}$ and $\cR^\prime_{6}$ contain the interactions with the combinations antisymmetric in $a$ and $b$. Therefore $\cR_5$ and $\cR^\prime_{5}$ provide altogether the flavour contractions of the SMEFT operator $\cO_1$. On the other hand, the interactions of $\cR^6$ and $\cR^\prime_{6}$ are described by a $d=8$ operator of the SMEFT Lagrangian.

The independent structures contained in the two remaining $RRLL$ operators, $\cR_7$ and $\cR_8$, read in the unitary gauge
\beq
\bar{u}^C_{R\alpha a}\,u_{R\beta b}\,\bar{d}^C_{L\gamma c}\,e_{L d}\varepsilon_{\alpha \beta\gamma}\,,\qquad
\bar{d}^C_{R\alpha a}\,d_{R\beta b}\,\bar{u}^C_{L\gamma c}\,\nu_{L d}\varepsilon_{\alpha \beta\gamma}\,,
\eeq
and are non-vanishing only for the combinations antisymmetric in $a$ and $b$. As a result, the number of independent flavour contractions for each of these operators is $N_f^3(N_f-1)/2$. 

The counting for the $Q_{L}Q_{L}Q_{R}L_{R}$ ($LLRR$) operators is not fully analogous to that of the $RRLL$ ones. The interactions in $\cR^9$ and $\cR^{10}$ are described by linear combinations of the operators $\cO_2$ and $\cO_5$ of the SMEFT Lagrangian, as in Eq.~(\ref{linear}). The number of their flavour contractions is $N_f^3(N_f+1)/2$ for each of them, in agreement with Ref.~\cite{Alonso:2014zka,Liao:2016qyd}. Finally, the counting of the flavour contractions of $\cR_{11}$ and $\cR_{12}$ is analogous to the one for their $RRLL$ counterparts, $\cR_7$ and $\cR_8$: $N_f^3(N_f-1)/2$.

As a result of the previous discussion, the number of flavour contractions can be summarised as follows:
\beq
\begin{aligned}
&\cR_1
\rightarrow 
N_f^2(2N_f^2+1)/3\\
&\cR_2
\rightarrow 
N_f^2(2N_f^2+1)/3\\
&\cR_3,\,\cR^\prime_3
\rightarrow 
N_f^4\\
&\cR_4,\,\cR^\prime_4
\rightarrow 
N_f^4\\
&\cR_5,\, \cR^\prime_5
\rightarrow 
N_f^4\\
&\cR_6,\,\cR^\prime_6
\rightarrow 
N_f^4
\end{aligned}
\qquad
\begin{aligned}
&\cR_7
&&\rightarrow 
N_f^3(N_f-1)/2\\
&\cR_8
&&\rightarrow 
N_f^3(N_f-1)/2\\
&\cR_9
&&\rightarrow 
N_f^3(N_f+1)/2\\
&\cR_{10} 
&&\rightarrow 
N_f^3(N_f+1)/2\\
&\cR_{11}
&&\rightarrow 
N_f^3(N_f-1)/2\\
&\cR_{12}
&&\rightarrow 
N_f^3(N_f-1)/2\,.\\
\end{aligned}      
\eeq 

\begin{center}
\rule[2mm]{2cm}{1pt}
\end{center}

This analysis completes previous studies on the HEFT Lagrangian, which received much attention in the last years for its relevance in collider searches. This paper provides, for the first time, the complete set of leading operators which are not invariant under baryon and lepton numbers, but do preserve $B-L$ combination. 

A detailed comparison with the SMEFT Lagrangian is also presented, pointing out a strategy to distinguish between the two approaches. Finally, the Hilbert series technique, which has recently undergone a revival of interest, has been adopted to discuss the number of flavour independent contractions for a generic number of fermion families.

\acknowledgments
We specially acknowledge discussions with Ilaria Brivio, Bel\'en Gavela, Elizabeth Jenkins and Aneesh Manohar. We also acknowledge Enrique Fern\'andez Mart\'inez for pointing out a typo in the text. S.S. is grateful to the Physics Department of the University of California, Berkeley, for hospitality during the completion of this work. L.M. and S.S. acknowledge partial financial support by the European Union through the FP7 ITN INVISIBLES (PITN-GA-2011-289442), by the Horizon2020 RISE InvisiblesPlus 690575, by CiCYT through the projects FPA2012-31880 and FPA2016-78645, and by the SpanishMINECO through the Centro de excelencia Severo Ochoa Program under grant SEV-2012-0249. The work of L.M. is supported by the Spanish MINECO through the ``Ram\'on y Cajal'' programme (RYC-2015-17173). The work of S.S. was supported through the grant BES-2013-066480 of the Spanish MICINN.


\providecommand{\href}[2]{#2}\begingroup\raggedright\endgroup

\end{document}